# Evolution of the band alignment at polar oxide interfaces

J. D. Burton[*] and E. Y. Tsymbal

*Department of Physics and Astronomy, Nebraska Center for Materials and Nanoscience,*
*University of Nebraska, Lincoln, Nebraska 68588-0299, USA*

First-principles calculations demonstrate the evolution of the band alignment at $La_{0.7}A_{0.3}MnO_3|La_{1-x}A_x|TiO_2|SrTiO_3(001)$ heterointerfaces, where $A$ = Ca, Sr, or Ba, as the interfacial A-site composition, $La_{1-x}A_x$, is varied from $x = 0.5$ to $x = 1.0$. This variation leads to a linear change of the $SrTiO_3$ valence band offset with respect to the Fermi level of the $La_{0.7}A_{0.3}MnO_3$ metal electrode and hence to a linear change of the Schottky barrier height at this interface. The effect arises due to electrostatic screening of the polar interface which alters the interfacial dipole and hence the electrostatic potential step at this interface. We find that both the $La_{0.7}A_{0.3}MnO_3$ and $SrTiO_3$ layers contribute to screening with both electronic and ionic screening being important for the change in the interface dipole. The results are in agreement with the recent experimental data.

The band alignment at metal/semiconductor and metal/insulator interfaces determines functional properties of such interfaces and thus is an important characteristic for applications. In particular, for a long time the physical mechanisms responsible for the formation of a Schottky barrier relevant to semiconductor electronics have been a subject of debate (see, e.g., Ref. [1]). Bardeen[2] showed that the presence of interface states due to crystal imperfections can account for the observed insensitivity of Fermi-level pinning to metal over-layer. Heine[3] argued that interface states are an intrinsic property that can be described as evanescent wave functions tunneling into the semiconductor band gap. Andrews and Phillips[4] ascribed the Fermi-level pinning to interfacial dipoles arising from polar bonds across the metal-semiconductor junction. Several models have been developed after these early researchers, such as models based on chemical bonding,[5,6] native defects[7] and metal induced gap states.[8,9]

More recently, the issue of the band alignment became critical for understanding properties of complex oxide heterostructures. Advances in oxide thin-film deposition techniques allow fabricating structures with atomic scale precision and have led to the discovery of new phenomena and properties at oxide heterointerfaces (see, e.g. Ref. [10] for a recent review). Elucidating the mechanisms responsible for determining the band alignment at oxide interfaces is important for engineering novel oxide-based devices. Compared to conventional semiconductors and metals, however, oxide heterointerfaces are far less understood. In particular, complex oxide materials are largely ionic in nature which affects significantly their interface properties. Different interface terminations, sometimes having a polar component, lead to a different charge transfer across them which determines the band alignment.[11,12] By being able to tailor the interface termination one can therefore manipulate barrier heights which are decisive for device applications.

In this work we employ first-principles calculations based density functional theory (DFT) to investigate the evolution of the band alignment at $La_{0.7}A_{0.3}MnO_3|La_{1-x}A_x|TiO_2|SrTiO_3(001)$ heterointerfaces, where $A$ = Ca, Sr, or Ba, as the interfacial composition, $La_{1-x}A_x$, is varied. The objective for this study is to demonstrate quantitatively a possibility to alter band offsets across oxide interfaces through interface doping. In addition to providing a model system for these studies, this and similar interfaces are of significant interest in magnetic tunnel junctions,[13,14] ferroelectric and multiferroic tunnel junctions[15-18], and electrically controlled interface magnetic structures.[19-21] We find that the variation of the interface composition leads to a linear change of the $SrTiO_3$ valence band offset with respect to the Fermi level of the $La_{0.7}A_{0.3}MnO_3$ metal electrode and hence to a linear change of the Schottky barrier height at this interface. The effect arises due to electrostatic screening of the polar interface which alters the interfacial dipole and hence the electrostatic potential step at this interface as the interfacial composition is varied. Our results are in qualitative agreement with the recent experimental data by Hikita et al.[11] who found that the fractional deposition of a $SrMnO_3$ unit cell at the $La_{0.7}Sr_{0.3}MnO_3/Nb:SrTiO_3$ interface led to a systematic increase in the n-type Schottky barrier height. Similar results were also recently reported by Minohara et al.[12]

Density-functional calculations are performed using the plane-wave pseudopotential code package Quantum-ESPRESSO.[22] The exchange-correlation functional was treated in the Perdew-Burke-Ernzerhof generalized gradient approximation (GGA).[23] The spin-polarized, self-consistent calculations were performed using an energy cutoff of 400 eV for the plane wave expansion and a 6×6×1 Monkhorst-Pack grid for k-point sampling. Atomic positions were converged until the Hellmann-Feynman forces on each atom became less than 20 meV/Å. Subsequent local density of states (LDOS) calculations were performed using a 20×20×1 Monkhorst-Pack k-point grid.

Fig. 1a shows the supercell which consists of a metal $La_{0.7}A_{0.3}MnO_3$ (LAMO) layer and an insulating $SrTiO_3$ (STO) layer stacked along the [001] direction of the conventional cubic perovskite cell. The metal layer consists of 9.5 unit cells of LAMO terminated on both ends by a $La_{1-x}A_xO$ atomic layer. The STO layer consists of 3.5 unit-cells and is terminated on both sides with $Ti-O_2$ atomic layers. Using a thicker layer of STO does not appreciably change the calculated band offsets or dipoles. We treat the $La_{1-x}A_x$ substitutional disorder within the virtual crystal approximation[24] where the perovskite A-site position is occupied by a pseudopotential generated for a fictitious atom with atomic number $57x + 56(1 - x)$ as in our previous work.[20] Technically this is closest to representing the Ba-doped manganite, and therefore the $x = 1.0$ situation corresponds to a pure BaO layer at the interface. The virtual-crystal pseudopotentials for the $La_{1-x}A_x$ atoms are created using Vanderbilt's Ultra-Soft-PseudoPotential generation code.[25, 26] The in-plane lattice constant is constrained to the calculated value for bulk cubic $SrTiO_3$, $a$ = 3.937Å, to simulate epitaxial growth on a $SrTiO_3$ substrate. This lattice constant is very close to the calculated lattice constant of cubic $La_{0.7}A_{0.3}MnO_3$ and therefore there is almost no strain induced in the LAMO layer which is found to be ferromagnetically ordered in all our calculations.

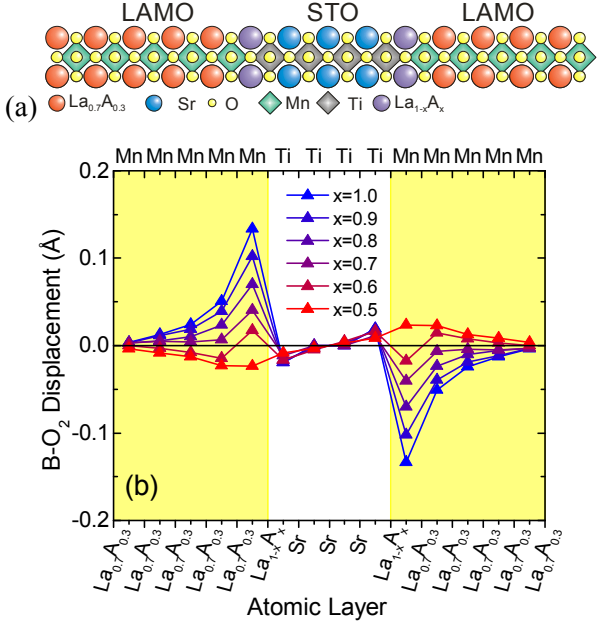

FIG 1. (a) Atomic structure of the supercell used in the calculations. (b) Intra-plane metal-oxygen displacements for various interface compositions. The points are Ba (Mn or Ti) - $O_2$ displacements along the [001] axis.

Fig. 1b shows the metal cation shift perpendicular to the planes with respect to the neighboring O atoms in the $Mn-O_2$ and $Ti-O_2$ monolayers for various interfacial compositions. There are also polar displacements in the $La_{1-x}A_x$-O and Sr-O atomic layers, but we do not plot them here. Clearly these polar displacements are largest in the LAMO monolayers near the interfaces. These shifts decrease and change sign as the interfacial A-site composition becomes more La-rich (i.e. smaller $x$).

To determine the band-alignment across the LAMO/STO interface we calculate the local density of states (LDOS) projected on to the central $SrTiO_3$ unit-cell in the heterostructure for all interface compositions. The results are plotted in Fig. 2, where the valence band maximum, $E_{VBM}$, is indicated as a solid vertical line. There is a systematic increase in the magnitude of the valence band offset, $E_{VO} = E_{VBM} - E_F$, as the interface composition $x$ decreases. The conduction band offset, $E_{CO}$, which determines the n-type Schottky barrier height, can be found from $E_{CO} = E_g + E_{VO}$, where $E_g$ = 1.82 eV is the theoretical band gap of STO, and increases systematically with $x$. Around $x$ = 0.5 and below we find that the conduction band minimum, $E_{CBM}$, falls below the Fermi level and electrons begin to occupy the conduction band states of STO. This unphysical situation is due to the underestimated DFT band-gap as compared to the experimental band gap of STO, $E_g$ = 3.2 eV. Therefore we limit our discussion to only those cases where this spurious behavior does not alter our results. In Fig. 3a we plot $E_{VO}$ for the whole range of interface compositions. We find a clear linear dependence which is consistent with the measurements of n-type Schottky barrier height reported by Hikita et al.[11]

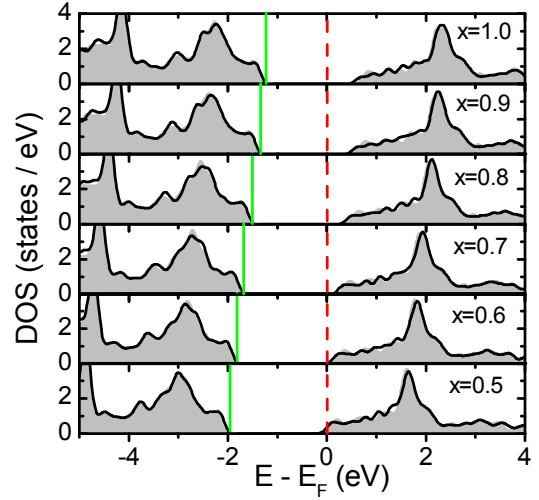

FIG. 2 LDOS projected on the central $SrTiO_3$ unit-cell in the LAMO/STO heterostructure for the series of interface compositions. The filled curves are for the majority-spin LDOS and the unfilled curves are for the minority-spin DOS. The vertical dashed line indicates the position of the Fermi level and the vertical solid lines indicate $E_{VBM}$.

In addition to the fully relaxed structure we also calculated the evolution of $E_{VO}$ in the case where the layer-by-layer polar displacements are completely suppressed. This "frozen-lattice" structure removes any contribution to $E_{VO}$ from lattice distortions and leaves only the electronic response. Fig. 3a shows $E_{VO}$ for $x$ = 0.7-1.0. It is seen that the slope is much larger than the fully relaxed case.

The dependence of $E_{VO}$ on the interface termination originates from the variation of the electrostatic dipole density, $D$, at the LAMO/STO interface with $x$. In the absence of any electrostatic interface dipole $E_{VO}$ can be extracted from separate bulk calculations of STO and LAMO by taking the difference between $E_F$ in LAMO and $E_{VBM}$ in STO calculated with respect to a reference potential. In LAMO the Fermi level, $E_F$, lies 9.10 eV above the average electrostatic potential energy, whereas in STO the $E_{VBM}$ lies 6.85 eV above. Therefore if the interface could be prepared in such a way that there was no electrostatic dipole, i.e. the electrostatic potential was aligned across the interface, then $E_{VO}^0$ = 2.25 eV.

In reality, however, there is charge rearrangement across this interface, resulting in the formation of the interface dipole layer $D$ and thus in the electrostatic potential energy step $eD/\varepsilon_0$. This potential energy step adds to the $E_{VO}^0$ resulting in the net valence band offset:

$$E_{VO} = E_{VO}^0 + eD/\varepsilon_0. \quad (1)$$

We calculate $D$ from the macroscopically averaged total (ionic + electronic) charge density profile, $\rho(z)$, where $z$ is the distance perpendicular to the plane from the center of the STO layer. The macroscopic average is obtained from the microscopic charge density, $\rho_{micro}(\mathbf{r})$, which includes the continuous charge distribution of all electrons, plus the delta-function-like charges of the nuclei.[27] We average $\rho_{micro}(\mathbf{r})$ over the plane parallel to the interface and then average along the z-axis using a window of width $a$ = 3.937Å. The resulting charge density profiles for both the fully relaxed and frozen lattice structures are plotted in Figs. 4a and 4b respectively for different interface compositions.

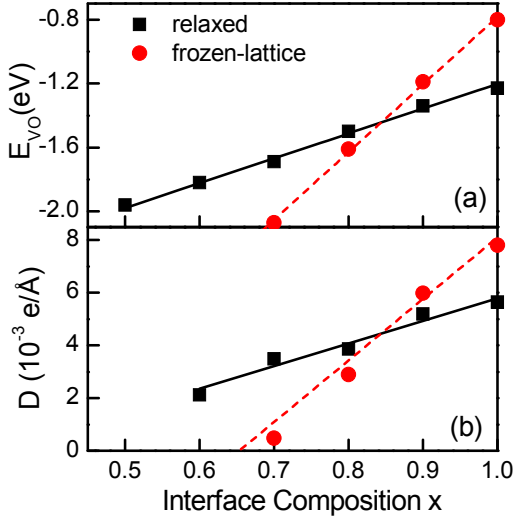

FIG. 3 Interface composition dependence of the valence band offset, $E_{VO}$, (a) and the interface dipole, $D$, (b). The squares are for the fully relaxed structure and the circles are for the frozen-lattice structure. The solid and dashed lines in (a) are least-squares linear fits to the relaxed and frozen-lattice data, respectively. The solid and dashed lines in (b) are the same as in (a) but translated to the corresponding $D$ values according to Eq. (1).

The interface dipole density is determined from $\rho(z)$ by

$$D = \int_{z_0}^{z_1} z\rho(z)dz, \quad (2)$$

where we limit the integration to a single supercell by choosing $z_0$ at the center of the STO (i.e. $z = 0$ in Fig. 4) and $z_1$ at the center of the LAMO layer ($z \approx 27.5$ Å in Fig. 4). There is an equal and opposite interface dipole at the other interface. The calculated interface dipole densities are plotted by dots in Fig. 3b for both the fully relaxed and frozen lattice structures. The lines in Fig. 3b are the same lines as in Fig. 3a but translated to the corresponding $D$ values using Eq. (1). The fact that these lines also fit the calculated values of $D$ confirms that the evolution of $E_{VO}$ with interface termination is determined solely by the change in $D$.

The predicted change in $D$ suggests that the changing interface composition $x$ leads to such a distribution charge across the interface that alters the magnitude of the interface dipole density. To comprehend this fact, first, we note that LAMO has a polar component at its interface. If we assign ionic charges to each site: $(La_{1-x}A_x)^{(3-x)+}$, $(La_{0.7}A_{0.3})^{2.7+}$, $Mn^{3.3+}$ and $O^{2-}$, then it appears that LAMO in fact consists of charged (001) planes reminiscent of a polar oxide insulator, e.g. $LaAlO_3$. Because of these charged planes there would be a net macroscopic interface charge density $\sigma = -e(x - 0.65)/a^2$. Since LAMO is a metal, however, free carriers rearrange to screen the divergent electric field due to this net interface charge. It is this charge screening which controls the interface dipole and therefore plays an important role in determining the band alignment.

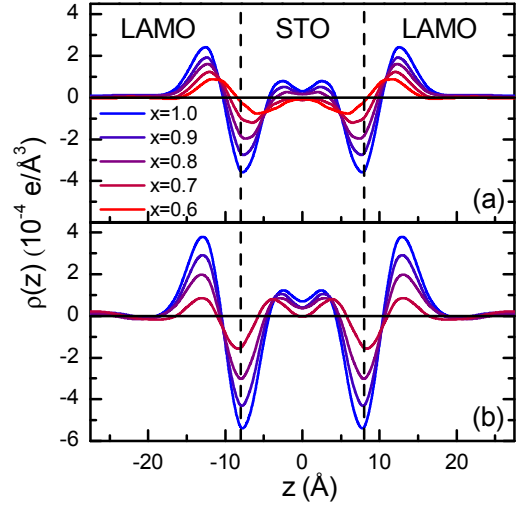

FIG. 4 Macroscopically averaged total charge density, $\rho(z)$, for the fully relaxed (a) and lattice-frozen (b) LAMO/STO heterostructure. The vertical dashed lines denote the interfacial $La_{1-x}A_x$ monolayer.

To see this we examine a simple model where we assume that the screening charge is entirely located within a semi-infinite LAMO metal layer with surface/interface at $z = 0$ and surface charge $\sigma$. The distribution of this screening charge is determined by the Thomas-Fermi screening length $\lambda$, so that the screening electron charge density exponentially decreases when moving from the interface to the bulk LAMO giving rise to the total free charge density

$$\rho_f(z) = \sigma\delta(z) - \frac{\sigma}{\lambda}e^{-z/\lambda}\begin{cases}0 & z<0 \\ 1 & z>0.\end{cases} \quad (3)$$

The electric displacement, $\mathbf{D}(z)$, associated with this free charge density is found by integrating $\nabla \cdot \mathbf{D}(z) = \rho_f$. From this one finds the electric field $\mathbf{E}(z) = \mathbf{D}(z)/\varepsilon$, where $\varepsilon$ is the dielectric constant of the LAMO due to the underlying ionic subsystem. The electrostatic potential step, $\Delta$, and therefore the interface dipole, can be calculated directly from $\mathbf{E}(z)$ by

$$\Delta\phi = \int_{-\infty}^{\infty} -E_z(z)dz = -\frac{\sigma\lambda}{\varepsilon}. \quad (4)$$

Hence the dipole density created by the charge distribution (3) is $-\sigma\lambda\varepsilon_0/\varepsilon$. Therefore, if for $x = 0$ the interface dipole density is $D_0$, then changing $x$ leads to the change in the interface dipole density according to $D = D_0 + ex\lambda\varepsilon_0/\varepsilon a^2$.

The increasing $D$ with $x$ is consistent with the results of calculations shown in Fig. 3b. In addition since the screening length, $\lambda$, is proportional to $\sqrt{\varepsilon}$ due to the ionic contribution to screening we therefore find that the slope of $D$ versus $x$ is inversely proportional to $\sqrt{\varepsilon}$. The fact that the slope in Fig. 3 is larger for the frozen-lattice system is consistent with the notion that $\varepsilon$ is increased by allowing the polar distortions seen in Fig. 1b to develop. These polar distortions are consistent with the lattice polarization found in our simple model: $\mathbf{P}(z) = \mathbf{D}(z)\,(\varepsilon - \varepsilon_0)/\varepsilon$. Assuming a Fermi-level density of states in the LAMO of 0.6/eV per unit-cell, we find from the fitted slopes in Fig. 3 that the dielectric constants are $\varepsilon = 32\varepsilon_0$ and $\varepsilon = 4.3\varepsilon_0$ for the relaxed and frozen lattice structures, respectively. In the frozen-lattice case the only contribution to $\varepsilon$ comes from the electronic deformation of the ionic cores, making $\varepsilon$ smaller than the relaxed case.

We note that the "ferrodistortive instability" reported by Pruneda et al.[28] at the bare surface of $La_{0.7}Sr_{0.3}MnO_3$ originates from precisely this screening of a polar surface charge. There a Mn-O$_2$ surface termination was assumed, which corresponds to a net negative surface charge. This gives rise to the reported off-centering where the Mn cations were shifted toward the surface relative to the O anions, consistent with the polarization found in our model above.

Our simple model obviously ignores the fact that the charge distribution at the interface is affected by the STO layer. First, it is clear from Fig. 1b that the polar distortions penetrate into the STO layer, giving rise to a bound polarization charge which contributes to the screening of the surface charge. Second, although STO is an insulator, metal-induced gap states[3] penetrate into the interior of STO and partly screen the electric field originating from the polar interface.[29] Both contributions to this non-vanishing charge within the STO are evident from Fig. 4 for both fully relaxed and frozen-lattice structures. In the latter case (Fig. 4b), for $x = 0.7$ the charge distribution at the interface is almost symmetric between metal and non-metal constituents resulting in a close to zero dipole moment. With increasing $x$ most of the screening change resides in the LAMO layer producing a large dipole moment and a strong change in $D$ with composition $x$. For the relaxed structure (Fig. 4a), however, for $x = 0.7$ the relaxation leads to the formation of a pronounced dipole layer with negative charge residing in STO and positive in LAMO. With increasing $x$ the ionic screening leads to a less prominent variation in magnitude of the dipole as compared to the unrelaxed structure. Thus, it is evident that both the LAMO and STO layers contribute to screening of the polar interface with both electronic and ionic screening being important for the formation of the dipole.

In conclusion, using first-principles DFT calculations we have explored the possibility to control the band alignment, and consequently the Schottky barrier height, at the (001) interface of two oxide materials $La_{0.7}A_{0.3}MnO_3$ (where $A$ = Ca, Sr, or Ba) and $SrTiO_3$ by changing the composition of the interfacial $La_{1-x}A_x$ monolayer. We found that the band offset changes linearly with $x$ solely due to the change in the electrostatic interface dipole density, $D$. The latter is determined by the screening of the electric field associated with the polar interface that occurs through electronic and ionic rearrangement within about 2 nm near the interface and involves both the $La_{0.7}A_{0.3}MnO_3$ and $SrTiO_3$ layers. Our results are in agreement with the available experimental data.[11,12] The predictions made have implications for the control of the band alignments across interfaces and thus may be important in the design of oxide heterostructures with new properties.

This work was supported by the National Science Foundation (Grant No. DMR-0906443), the Nanoelectronics Research Initiative of the Semiconductor Research Corporation, the Materials Research Science and Engineering Center (NSF Grant No. DMR-0820521), and the Nebraska Research Initiative. Computations were performed utilizing the Research Computing Facility of the University of Nebraska-Lincoln and the Center for Nanophase Materials Sciences at Oak Ridge National Laboratory.

* jdburton1@gmail.com

[1] S. M. Sze, Physics of Semiconductor Devices (Wiley, New York, 1981).
[2] J. Bardeen, Phys. Rev. 71, 717 (1947).
[3] V. Heine, Phys. Rev. 138, A1689 (1965).
[4] J. M. Andrews and J. C. Phillips, Phys. Rev. Lett. 35, 56 (1975).
[5] L. J. Brillson, Phys. Rev. Lett. 40, 260 (1978).
[6] R. T. Tung, Phys. Rev. Lett. 84, 6078 (2000).
[7] W. E. Spicer, I. Lindau, P. Skeath, and C. Y. Su, Journal of Vacuum Science and Technology 17, 1019 (1980).
[8] S. G. Louie, J. R. Chelikowsky, and M. L. Cohen, Phys. Rev. B 15, 2154 (1977).
[9] J. Tersoff, Phys. Rev. B 32, 6968 (1985).
[10] J. Mannhart and D. G. Schlom, Science 327, 1607 (2010).
[11] Y. Hikita, M. Nishikawa, T. Yajima, and H. Y. Hwang, Phys. Rev. B 79, 073101 (2009).
[12] M. Minohara, R. Yasuhara, H. Kumigashira, and M. Oshima, Phys. Rev. B 81, 235322 (2010).
[13] J. M. De Teresa, et al., Science 286, 507 (1999).
[14] E. Y. Tsymbal, K. D. Belashchenko, J. P. Velev, S. S. Jaswal, M. van Schilfgaarde, I. I. Oleynik, and D. A. Stewart, Progress in Materials Science 52, 401 (2007).
[15] M. Y. Zhuravlev, R. F. Sabirianov, S. S. Jaswal, and E. Y. Tsymbal, Phys. Rev. Lett. 94, 246802 (2005).
[16] E. Y. Tsymbal and H. Kohlstedt, Science 313, 181 (2006).
[17] J. P. Velev, C.-G. Duan, J. D. Burton, A. Smogunov, M. K. Niranjan, E. Tosatti, S. S. Jaswal, and E. Y. Tsymbal, Nano Lett. 9, 427 (2009).
[18] V. Garcia, et al., Science 327, 1106 (2010).
[19] H. J. A. Molegraaf, J. Hoffman, C. A. F. Vaz, S. Gariglio, D. v. d. Marel, C. H. Ahn, and J.-M. Triscone, Advanced Materials 21, 3470 (2009).
[20] J. D. Burton and E. Y. Tsymbal, Phys. Rev. B 80, 174406 (2009).
[21] C. A. F. Vaz, J. Hoffman, Y. Segal, J. W. Reiner, R. D. Grober, Z. Zhang, C. H. Ahn, and F. J. Walker, Phys. Rev. Lett. 104, 127202 (2010).
[22] P. Giannozzi, et al., J. Phys.: Cond. Mat. 21, 395502 (2009).
[23] J. P. Perdew, K. Burke, and M. Ernzerhof, Phys. Rev. Lett. 77, 3865 (1996).
[24] L. Nordheim, Annalen der Physik 401, 607 (1931).
[25] D. Vanderbilt, Phys. Rev. B 41, 7892 (1990).
[26] D. Vanderbilt, http://www.physics.rutgers.edu/~dhv/uspp/.
[27] M. Peressi, N. Binggeli, and A. Baldereschi, J. Phys. D 31, 1273 (1998).
[28] J. M. Pruneda, V. Ferrari, R. Rurali, P. B. Littlewood, N. A. Spaldin, and E. Artacho, Phys. Rev. Lett. 99, 226101 (2007).
[29] K. Janicka, J. P. Velev, and E. Y. Tsymbal, Phys. Rev. Lett. 102, 106803 (2009).